# Numerical Simulation of Lead-Free Absorbers in 2D Dion–Jacobson Phase Perovskite Solar Cells Using SCAPS-1D: Towards 41% Efficiency

Md. Meraz Hasan[a], Pallab Chakraborty[a], Fahim Tanvir[a], Subah Tahsin[a], Mostafizur Rahaman[a]

[a]Department of Energy Science and Engineering, Khulna University of Engineering & Technology (KUET), Khulna 9203, Bangladesh

**Abstract**

With the rapid advancement of photovoltaic science, there has been an increasing focus on the development of environment-friendly and structurally advanced perovskite solar cells (PSCs). In this context, this study investigates an architectural configuration employing 2D Dion–Jacobson phase perovskites as both electron and hole transport layers within a 2D/absorber/2D structure. The primary objective is to identify optimal absorber materials and enhance the overall efficiency of the device. While the reduction of lead content remains a significant challenge in PSC development, the present work focuses on the evaluation of seven lead-free absorber materials: $MASnBr_3$, $Sr_3PI_3$/$Sr_3SbI_3$, p-$CuBi_2O_4$, p-Si, $CH_3NH_3SnI_3$, $Sb_2Se_3$, and CZTSSe. These materials were assessed in the context of an FTO/$PeDAMA_8Pb_6I_{19}$/IDL1/absorber/IDL2/$PeDAMA_2Pb_3I_{10}$/C architecture utilizing SCAPS-1D simulation software. The study includes a comprehensive analysis of band alignment, pre-optimization screening, and performance optimization through the adjustment of absorber thickness, doping levels, and defect densities. Additionally, the temperature sensitivity and the substitution of the FTO layer with ITO, IZO, and MZO were also investigated. The simulation results indicated that $Sb_2Se_3$ and CZTSSe achieved the highest efficiencies of 41.00% and 41.19%, respectively. Furthermore, MZO was identified as a strong candidate for replacing FTO, maintaining consistent performance across all absorber types analyzed. Overall, this study provides a comparative framework for the material selection within layered PSC architectures and significantly contributes to the advancement of stable, efficient, and lead-free photovoltaic technologies.

Corresponding author: Mostafizur Rahaman. E-mail address: mostafiz@ese.kuet.ac.bd

1. Introduction

To address the energy crisis driven by rapid urbanization and industrialization, solar energy has been considered a predominant source due to its abundant availability and minimal environmental impact. It offers a sustainable solution to global energy demands while simultaneously diminishing reliance on fossil fuels [1,2]. Within the realm of solar technologies, Silicon (Si) solar cells dominate the photovoltaic (PV) market due to their remarkable efficiency, stability, and cost-effectiveness [3,4]. However, these cells encounter several inherent limitations. One major constraint is the theoretical efficiency boundary of approximately 29.4%, established by the Shockley–Queisser limit, with leading experimental cells achieving a maximum efficiency of 27.4% [5,6]. As an indirect bandgap material, silicon's capacity to absorb light is limited, particularly in the infrared spectrum, thus necessitating the employment of comparatively thick wafers and advanced light-trapping technologies [7]. Additional efficiency constraints arise from surface and interface recombination [4,8], parasitic absorption within contact layers [9], and performance degradation at high temperatures [10]. In response to these challenges, perovskite solar cells (PSCs) have revolutionized photovoltaics, achieving unprecedented efficiency leaps from 3.8% in 2009 to over 25% in more recent developments. This rapid progression positions PSCs as serious contenders to traditional silicon-based technologies [11,12]. Nevertheless, despite their high power conversion efficiencies, conventional perovskite solar cells (PSCs) face several challenges that hinder their commercialization. A primary limitation of perovskite solar cells is their inadequate environmental stability, as exposure to moisture and ultraviolet (UV) radiation causes deterioration of the perovskite layers. This results in device lifetimes that often fall below 10,000 hours, well short of the 25-year benchmark set by silicon-based alternatives [13,14]. Additionally, PSCs are afflicted by surface and interface defects, charge extraction losses, and ion migration, which collectively contribute to diminished long-term performance [15–20].

Among the proposed architectural innovations of PSCs, the emerging configuration of ETL (2D)/absorber/HTL (2D) effectively addresses these challenges. This design integrates an absorber with 2D Dion–Jacobson (DJ) phase perovskite layers at both interfaces. Existing studies predominantly focus on three-dimensional (3D) materials as absorbers, which provide excellent charge transport and robust light absorption. Concurrently, the outer 2D perovskite layers function as passivating and protective barriers. This stacked architecture significantly mitigates interfacial challenges, encompassing non-radiative recombination, ion migration, and environmental degradation [15,21–23].

The role of the inner 3D absorber layer is pivotal for efficient photogeneration and carrier transport. Commonly used 3D absorbers include $MASnBr_3$ [24–26], $CH_3NH_3SnI_3$ [27–29], and CZTSSe [30–32]. These materials are characterized by their continuous crystalline lattices, which facilitate isotropic charge transport. Furthermore, $Sr_3PI_3$ and $Sr_3SbI_3$, identified as 3D inorganic perovskite materials that are employed within tandem PSCs as top and bottom absorbers, respectively [33–35]. Other noteworthy 3D materials include p-$CuBi_2O_4$, a metal oxide with strong visible light absorption and thermal stability [36]. Additionally, p-type silicon (p-Si) is widely used in tandem architectures due to its mechanical strength and electronic stability. In contrast, $Sb_2Se_3$ exhibits a one-dimensional (1D) ribbon-like crystal structure, which enables anisotropic carrier transport and makes it suitable for use as a bottom absorber in tandem designs [37–39]. The outer 2D perovskite layers, such as $PeDAMA_8Pb_6I_{19}$ and $PeDAMA_2Pb_3I_{10}$, act as advanced electron and hole transport layers (ETLs and HTLs) in PSCs. Their hydrophobicity, thermal stability, and ability to mitigate surface defects and ion migration enhance operational stability and minimize hysteresis [16,19,20,22]. Furthermore, these layers provide favorable energy level alignment, which reduces interfacial resistance and facilitates higher open-circuit voltage and fill factor [40–42]. In comparison to conventional transport layers like $TiO_2$, $SnO_2$, Spiro-OMeTAD, and PEDOT:PSS, 2D perovskites offer superior moisture resistance and interface engineering capabilities, contributing to power conversion efficiencies exceeding 26% [43–46].

To evaluate the performance of this architecture, we conducted SCAPS-1D simulations for both a 2D/3D/2D configuration and a 2D/1D/2D configuration. The study incorporated six distinct absorber materials: $MASnBr_3$, $Sr_3PI_3$/$Sr_3SbI_3$, p-$CuBi_2O_4$, p-Si, $CH_3NH_3SnI_3$, and CZTSSe for the 2D/3D/2D arrangement, while employing $Sb_2Se_3$ as the absorber for the 2D/1D/2D setup. The comprehensive structural configuration under analysis was as follows: FTO/$PeDAMA_8Pb_6I_{19}$/IDL1/absorber/IDL2/$PeDAMA_2Pb_3I_{10}$/C.

The front contact layer in that architecture, fluorine-doped tin oxide (FTO), remains the most commonly used transparent conducting oxide (TCO) due to its high optical transparency, good electrical conductivity, chemical stability, and recyclability [47–50]. Its typical work function of approximately 4.4 eV facilitates effective carrier extraction at the front interface [51,52]. Alternative TCOs like indium tin oxide (ITO), indium zinc oxide (IZO), and magnesium zinc oxide (MZO) have also shown promise. ITO has demonstrated strong potential in n-i-p structured PSCs (ITO/SnO$_2$/MAPbI$_3$/spiro-OMeTAD/Au), achieving simulated power conversion efficiencies of around 24%. In contrast, IZO has been effectively integrated as front contacts in bifacial PSCs, reaching efficiencies as high as 26.03% [53]. MZO has also been explored for its tunable optical and electrical properties, thereby providing additional flexibility for device optimization [53].

At the opposite terminal, carbon (C) has been employed as the back electrode due to its exceptional environmental stability, cost-effectiveness, and compatibility with scalable deposition techniques such as printing and spray coating. In contrast to noble metals, such as gold (Au) and silver (Ag), carbon exhibits remarkable resistance to moisture and oxygen, thus enhancing device stability even without encapsulation [54,55]. Additionally, advanced forms such as graphene and carbon nanotubes (CNTs) provide favorable energy level alignment and efficient hole extraction, thereby further supporting performance optimization [56].

At the interfacial level, Interface Defect Layers (IDL1 and IDL2) play a crucial role in determining the efficiency of photovoltaic devices. IDL1, positioned between the electron transport layer (ETL) and the absorber, predominantly influences charge injection and recombination dynamics. Conversely, IDL2 is located at the absorber/hole transport layer (HTL) interface and significantly affects the open-circuit voltage ($V_{oc}$) by controlling non-radiative recombination processes [57]. Elevated defect densities exceeding $10^{15}$ cm$^{-3}$ at either interface can serve as recombination centers, leading to a reduction in $J_{sc}$, $V_{oc}$, and FF. This highlights the importance of effective interface passivation strategies, particularly through the incorporation of 2D perovskites, in enhancing device performance [58,59].

Within this simulation framework, recent studies have highlighted the chosen lead-free absorber materials with strong potential for high efficiency. Methylammonium tin bromide (MASnBr$_3$) is gaining traction due to its excellent visible light absorption and stable electronic structure. Single-junction solar cells incorporating this material have demonstrated efficiencies of up to 34.52%, while tandem configurations, when combined with $Cs_2TiBr_6$, approach efficiencies of 33% [26,60,61]. Interface engineering has also played a pivotal role in enhancing device performance; materials such as graphene and 3C–SiC have facilitated efficiencies exceeding 31% by improving charge extraction mechanisms [62]. Similarly, the $Sr_3PI_3/Sr_3SbI_3$ tandem system benefits from optimal bandgap alignment, achieving power conversion efficiencies (PCE) of up to 34.13% with strong current outputs and minimized recombination losses [63,64]. Additionally, copper bismuth oxide (p-CuBi2O4) offers a promising alternative to traditional solar materials, with simulations demonstrating over 25% efficiency when combined with transport layers like ZnO and CuSCN. However, achieving this efficiency requires precise optimization of the absorber thickness and band alignment [65,66]. In a different class, crystalline p-type silicon (p-Si) has established itself as a benchmark for hybrid tandem designs, yielding over 25% PCE in simulations through careful interface management and defect control [67,68]. Building on this foundation, simulated tandem configurations integrated with perovskite top cells indicate potential efficiencies exceeding 35% [68]. However, challenges remain, particularly in mitigating recombination at ETL/perovskite interfaces. Methylammonium tin iodide ($CH_3NH_3SnI_3$) has also garnered attention due to its lead-free nature and high carrier mobility. Such a device is $ZnS/CH_3NH_3SnI_3/MoO_3$, achieving an efficiency of up to 32.57% [69]. However, the practical application of this material is impeded by notable thermal instability stemming from the oxidation of $Sn^{2+}$ to $Sn^{4+}$ under ambient conditions [70]. Hence, achieving optimal performance necessitates refined doping strategies and effective defect passivation techniques [69]. Antimony selenide ($Sb_2Se_3$), with its ideal bandgap and low toxicity, has achieved a notable efficiency of 30% in single-layer devices with a thickness of 800 nm. Maintaining defect densities below $10^{15}$ cm$^{-3}$ is critical to ensure optimal device performance [71]. Lastly, CZTSSe continues to draw attention for its high theoretical efficiency and strong thermal stability. In combination with $CsPbI_3$, it achieves simulated efficiencies up to 37.35% when absorber thickness (500 nm–1.4 μm), defect density, and contact work function are appropriately optimized [30,72]. Although CZTSSe demonstrates thermal stability at 300 K for practical deployment, its efficiency tends to decrease with rising temperatures [73]. However, bandgap mismatches and structural incompatibilities in non-perovskite absorbers (e.g., $Sb_2Se_3$,

CZTSSe) limit their integration into perovskite-dominated stacks [74,75]. Despite notable advancements in this field, comprehensive comparative studies of emerging absorbers within unified architectural frameworks remain scarce, resulting in fragmented insights and a need for systematic evaluation [76,77].

Several critical gaps hinder this progress. First, there is a lack of studies on 2D/3D/2D stacked structures specifically involving the FTO/PeDAMA$_8$Pb$_6$I$_{19}$/IDL1/absorber/IDL2/PeDAMA$_2$Pb$_3$I$_{10}$/C configuration. Second, the performance of 1D absorbers, such as Sb$_2$Se$_3$, in this layered structure remains underexplored, as most research has focused on 3D absorber materials. Third, direct systematic comparisons of these seven lead-free absorber materials within the same architecture have not been thoroughly conducted. Fourth, the suitability of front contacts such as ITO, IZO, and MZO in this configuration has not yet been evaluated. These gaps have impeded the development of efficient, stable, and environmentally friendly PSCs.

This study bridges these gaps by simulating seven lead-free absorbers (MASnBr$_3$, Sr$_3$PI$_3$/Sr$_3$SbI$_3$, p-CuBi$_2$O$_4$, p-Si, CH$_3$NH$_3$SnI$_3$, Sb$_2$Se$_3$, and CZTSSe) of the proposed architecture to demonstrate pathways to eco-friendly PSCs without compromising performance. The initial performance evaluation (band diagrams, current density-voltage characteristics, and spectral responses) was conducted for each absorber. We optimized the device through a balanced trade-off between efficiency and stability. Additionally, we investigated the suitability of ITO, IZO, and MZO as alternatives to FTO as front electrodes, using the optimized parameters for each absorber. The simulation results reveal a highly efficient and stable absorber material, as well as a framework for selecting absorbers in similar configurations. Furthermore, the study identifies the best-performing front contact combination in the specified architecture. Nonetheless, our findings contribute to the transition from silicon to perovskite-based photovoltaics for advancing highly efficient and durable solar technologies.

## 2. Methodology

The photovoltaic performance of the multilayered perovskite solar cell was investigated using SCAPS-1D software version 3.3.09. **Fig. 1(a)** presents the proposed device architecture, which comprises FTO as the front contact, PeDAMA$_8$Pb$_6$I$_{19}$ as the electron transport layer (ETL), one of the seven materials such as MASnBr$_3$, Sr$_3$PI$_3$/Sr$_3$SbI$_3$, p-CuBi$_2$O$_4$, p-Si CH$_3$NH$_3$SnI$_3$, Sb$_2$Se$_3$ or CZTSSe as the absorber layer, PeDAMA$_2$Pb$_3$I$_{10}$ as the hole transport layer (HTL), and carbon (C) as the back electrode. The interface defect layers IDL1 and IDL2 were integrated between the absorber and ETL/HTL to incorporate the interfacial recombination effects. The energy-band alignment across the various layers within this architecture is illustrated in **Fig. 1(b).**

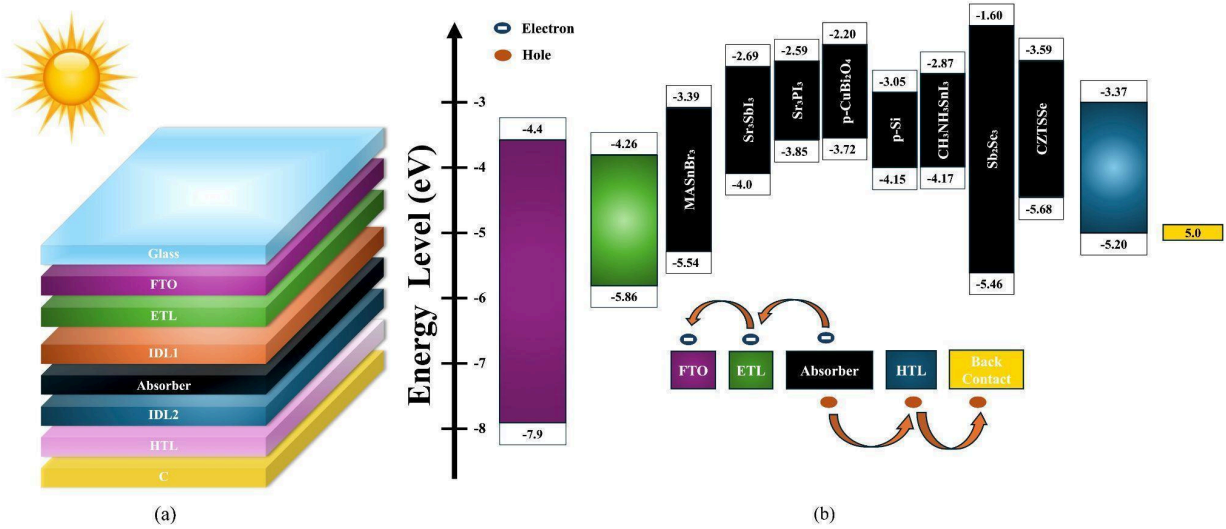

**Fig. 1** (a) Structural layout of the device layers; (b) Configured energy band structure.

The numerical analysis of the solar cell was carried out by modeling it using coupled ordinary differential equations. SCAPS-1D was employed to solve these equations and simulate charge carrier dynamics under steady-state conditions [78,79]. In this approach, Poisson's equation is used to determine the electric potential and internal electric field by relating the spatial charge density to potential variations, as shown in **Eq. (1)**.

$$-\frac{\partial}{\partial x}\left(\varepsilon(x)\frac{\partial V}{\partial x}\right) = q\left[p(x)n(x)N_D^+(x) - N_A^-(x) + P_t(x) - n_t(x)\right] \quad (1)$$

Here, $q$ is the electronic charge, $\varepsilon$ the dielectric constant, $V$ the potential, $p(x)$ and $n(x)$ are the free hole and electron concentrations, $N_D^+(x)$ and $N_A^-(x)$ are the donor and acceptor densities, respectively, and $P_t(x)$, $n_t(x)$ are the hole and electron trap concentrations, respectively. Carrier transport is described by the drift-diffusion equations formulated in **Eqs. (2a)** and **(2b)**.

$$J_n = q\mu_n pE + qDn\frac{dn}{dx} \qquad (2a)$$

$$J_p = q\mu_p pE + qDp\frac{dp}{dx} \qquad (2b)$$

where, $\mu_n$ and $\mu_p$ are electron/hole mobilities, $Dn$ and $Dp$ are diffusion coefficients, and $E$ is the electric field. The carrier continuity equations, as expressed in **Eqs. (3a)** and **(3b)**, account for generation G, recombination R, and the divergence of carrier flux.

$$\frac{\partial n}{\partial t} = \frac{1}{q}\frac{\partial J_n}{\partial x} + G - R \qquad (3a)$$

$$\frac{\partial p}{\partial t} = -\frac{1}{q}\frac{\partial J_p}{\partial x} + G - R \qquad (3b)$$

These equations collectively enable the self-consistent simulation of charge transport and device performance. Material properties such as the thickness ($d$), bandgap ($E_g$), electron affinity ($\chi$), dielectric constant ($\varepsilon$), carrier mobility (n, p), effective density of states ($N_c$, $N_v$), doping concentrations ($N_d$, $N_a$), and defect densities ($N_t$) are listed in **Table 1**. **Table 2** summarizes the initial parameters of seven absorber candidates: $MASnBr_3$, $Sr_3PI_3/Sr_3SbI_3$, p-$CuBi_2O_4$, p-Si, $CH_3NH_3SnI_3$, $Sb_2Se_3$, and CZTSSe. The defect characteristics, including the energy levels and capture cross sections, are detailed in **Table 3**. The device was illuminated under the AM1.5G spectrum at 1000 W/m² at 300 K. Light irradiation was applied from the right side with no neutral-density filter. The series and shunt resistances were turned off to isolate intrinsic material effects. Voltage bias was applied at the left contact, with the current referenced as a consumer. The layer thicknesses were varied to minimize recombination losses. The doping concentrations ($N_a$ and $N_d$) were optimized to balance the carrier transport. The defect densities ($N_t$) were adjusted to suppress non-radiative recombination. The temperature effects were analyzed to assess thermal stability. Performance parameters, such as open-circuit voltage ($V_{oc}$), short-circuit current ($J_{sc}$), fill factor (FF), and power conversion efficiency (PCE), were evaluated to assess the suitability of the absorber.

**Table 1** Basic parameters of the initial structure of the cell [80].

| Parameters | Transparent Conducting Substrate | Electron Transport Layer (ETL) | Interface Defect Layer (1) | Absorbers | Interface Defect Layer (2) | Hole Transport Layer (HTL) | Back Electrode |
|---|---|---|---|---|---|---|---|
|  | FTO | PeDAMA$_5$Pb$_6$I$_{19}$ | IDL1 |  | IDL2 | PeDAMA$_2$Pb$_3$I$_{10}$ | C |
| Thickness (nm) | 50 | 50 | 2 |  | 10 | 100 |  |
| Band gap $E_g$/(eV) | 3.5 | 1.6 | 1.6 |  | 1.3 | 1.83 |  |
| Electron affinity $\chi$/(eV) | 4 | 3.98 | 3.98 |  | 3.6 | 3.15 |  |
| Dielectric Permittivity $\varepsilon_r$ | 9 | 25 | 25 |  | 9.93 | 25 |  |
| CB effective density of states $N_c$/(cm$^{-3}$) | 2.20E+18 | 7.50E+17 | 7.50E+17 |  | 7.50E+17 | 7.50E+17 |  |
| VB effective density of states $N_v$/(cm$^{-3}$) | 1.80E+19 | 1.80E+18 | 1.80E+18 | MASnBr$_3$, Sr$_3$PI$_3$/Sr$_3$SbI$_3$, p-CuBi$_2$O$_4$, p-Si, CH$_3$NH$_3$SnI$_3$, Sb$_2$Se$_3$, CZTSSe | 1.80E+18 | 1.80E+18 |  |
| Electron thermal velocity (cm/s) | 1.00E+07 | 1.00E+07 | 1.00E+07 |  | 1.00E+07 | 1.00E+07 | Work function (eV)=5 |
| Hole thermal velocity (cm/s) | 1.00E+07 | 1.00E+07 | 1.00E+07 |  | 1.00E+07 | 1.00E+07 |  |
| Electron mobility $\mu_n$/(cm$^2$/Vs) | 20 | 1.4 | 1.4 |  | 1.4 | 1.4 |  |
| Hole mobility $\mu_p$/(cm$^2$/Vs) | 10 | 0.3 | 0.3 |  | 0.3 | 0.3 |  |
| Donor doping density $N_D$/(cm$^{-3}$) | 2.00E+19 | 1.00E+18 | 0 |  | 0 | 0 |  |
| Acceptor doping density $N_A$/(cm$^{-3}$) | 0 | 0 | 0 |  | 0 | 1.00E+15 |  |
| Defect density $N_t$/(cm$^{-3}$) | 1.00E+15 | 1.00E+15 | 1.00E+16 |  | 1.00E+20 | 1.00E+16 |  |

Table 2 Basic parameters of the absorber layer of the cell [26,30,63,81–84].

| Absorber | $MASnBr_3$ | $Sr_3SbI_3$ | $Sr_3PI_3$ | $p-CuBi_2O_4$ | p-Si | $CH_3NH_3SnI_3$ | $Sb_2Se_3$ | (CZTSSe) |
|---|---|---|---|---|---|---|---|---|
| Absorber layer type | Single | Double | | Single | Single | Single | Single | Single |
| Thickness (μm) | 0.5 | 0.5 | 0.5 | 2 | 2 | 1 | 0.6 | 1.4 |
| Band gap $E_g$/(eV) | 1.3 | 1.307 | 1.258 | 1.5 | 1.14 | 1.3 | 1.06 | 1.15 |
| Electron affinity χ/(eV) | 4.17 | 4 | 3.85 | 3.72 | 4.15 | 4.17 | 4.35 | 4.1 |
| Dielectric Permittivity $\varepsilon_r$ | 10 | 5.4 | 5.9 | 34 | 11.9 | 10 | 19 | 13.6 |
| CB effective density of states $N_c$/(cm$^{-3}$) | 2.20E+18 | 1.20E+19 | 1.70E+19 | 1.20E+19 | 2.80E+19 | 1.00E+18 | 1.00E+18 | 2.20E+18 |
| VB effective density of states $N_v$/(cm$^{-3}$) | 1.80E+18 | 2.40E+19 | 1.03E+19 | 5.00E+19 | 2.65E+19 | 1.00E+19 | 1.80E+20 | 1.80E+19 |
| Electron thermal velocity (cm/s) | 1.00E+07 | | | | | | | |
| Hole thermal velocity (cm/s) | 1.00E+07 | | | | | | | |
| Electron mobility $\mu_n$/(cm$^2$/Vs) | 1.60E+00 | 50 | 20 | 1.10E−03 | 1.45E+03 | 1.6 | 10 | 100 |
| Hole mobility $\mu_p$/(cm$^2$/Vs) | 1.60E+00 | 50 | 25 | 1.20E−03 | 5.00E+02 | 1.6 | 1 | 25 |
| Donor doping density $N_D$/(cm$^{-3}$) | 1.00E+15 | 0 | 0 | 0 | 0 | 0 | 0 | 0 |
| Acceptor doping density $N_A$/(cm$^{-3}$) | 1.00E+18 | 1.00E+18 | 1.00E+18 | 3.70E+18 | 1.00E+18 | 1.00E+17 | 2.00E+14 | 1.00E+18 |

**Table 3** Initial defect parameters settings of the absorbers and interface used in $Sr_3PI_3/Sr_3SbI_3$ of the solar cell [30,63,81,83–86].

| Defect parameters | MASnBr$_3$ | Interface ($Sr_3PI_3/Sr_3SbI_3$) | p-CuBi$_2$O$_4$ | p-Si | CH$_3$NH$_3$SnI$_3$ | Sb$_2$Se$_3$ | CZTSSe |
|---|---|---|---|---|---|---|---|
| Defect type | Neutral | Neutral | Neutral | Single Donor | Neutral | Neutral | Neutral |
| Capture cross section for electrons (cm$^2$) | 1.00E-15 | 1.00E-19 | 1.00E-15 | 1.00E-15 | 1.00E-15 | 1.00E-15 | 1.00E-15 |
| Capture cross section for holes (cm$^2$) | 1.00E-15 | 1.00E-19 | 1.00E-15 | 1.00E-15 | 1.00E-15 | 1.00E-15 | 1.00E-15 |
| Type of energetic distribution | Single | Single | Single | Gaussian | Gaussian | Single | Single |
| Reference for defect energy level | Above E$_v$ | Above E$_v$ | Above E$_v$ | Above E$_v$ | Above E$_v$ | Above E$_v$ | Above E$_v$ |
| Energy level with respect to E$_v$ (eV) | 0.6 | 0.06 | 0.6 | 0.6 | 0.6 | 0.6 | 0.6 |
| Characteristic Energy (eV) | – | – | – | 0.1 | – | – | – |
| Total defect density (cm$^{-3}$) | 1.00E+14 | 1.00E+10 | 1.00E+10 | 2.00E+14 | 1.00E+14 | 1.00E+10 | 1.00E+12 |

## 3. Result and Discussion

### 3.1. Band Diagram Analysis

To understand the charge transport behavior and interfacial dynamics within the photovoltaic device, it is essential to first assess the energy band alignments across the structure. **Fig. 2(a-g)** provides important insights into the electronic properties of seven absorbers within the FTO/ETL/IDL1/absorber/IDL2/HTL/C architecture. The diagrams highlight distinct band alignment patterns, with absorber-specific $E_c$-$E_v$ gaps predominantly ranging from approximately 1.3 to 1.5 eV. Notably, p-$CuBi_2O_4$ **(Fig. 2c)** exhibits wider band gaps that contrast sharply with narrow-gap p-Si **(Fig. 2d)**. Fermi levels are positioned near the conduction band edge ($E_c$) or the valence band edge ($E_v$), reflecting the type of doping; for instant, p-$CuBi_2O_4$ has its Fermi level ($F_p$) close to $E_v$, which enhances hole extraction. Conversely, the abrupt band bending observed at the ETL/absorber interfaces in $Sr_3PI_3$/$Sr_3SbI_3$ **(Fig. 2b)** and $Sb_2Se_3$ **(Fig. 2f)** indicates potential recombination risks. In contrast, $MASnBr_3$ **(Fig. 2a)** displays minimal band gradients, suggesting uniform carrier transport. Furthermore, $Sb_2Se_3$ exhibits a unique, gradual shift in the valence band that is not observed in other materials. These variations underscore the significance of interfacial energetics in determining charge separation and establishing performance limits before optimization. The material-specific characteristics, such as thickness-dependent bending and $E_c$/$E_v$ offsets, highlight the necessity for tailored designs to minimize losses and optimize photovoltaic response.

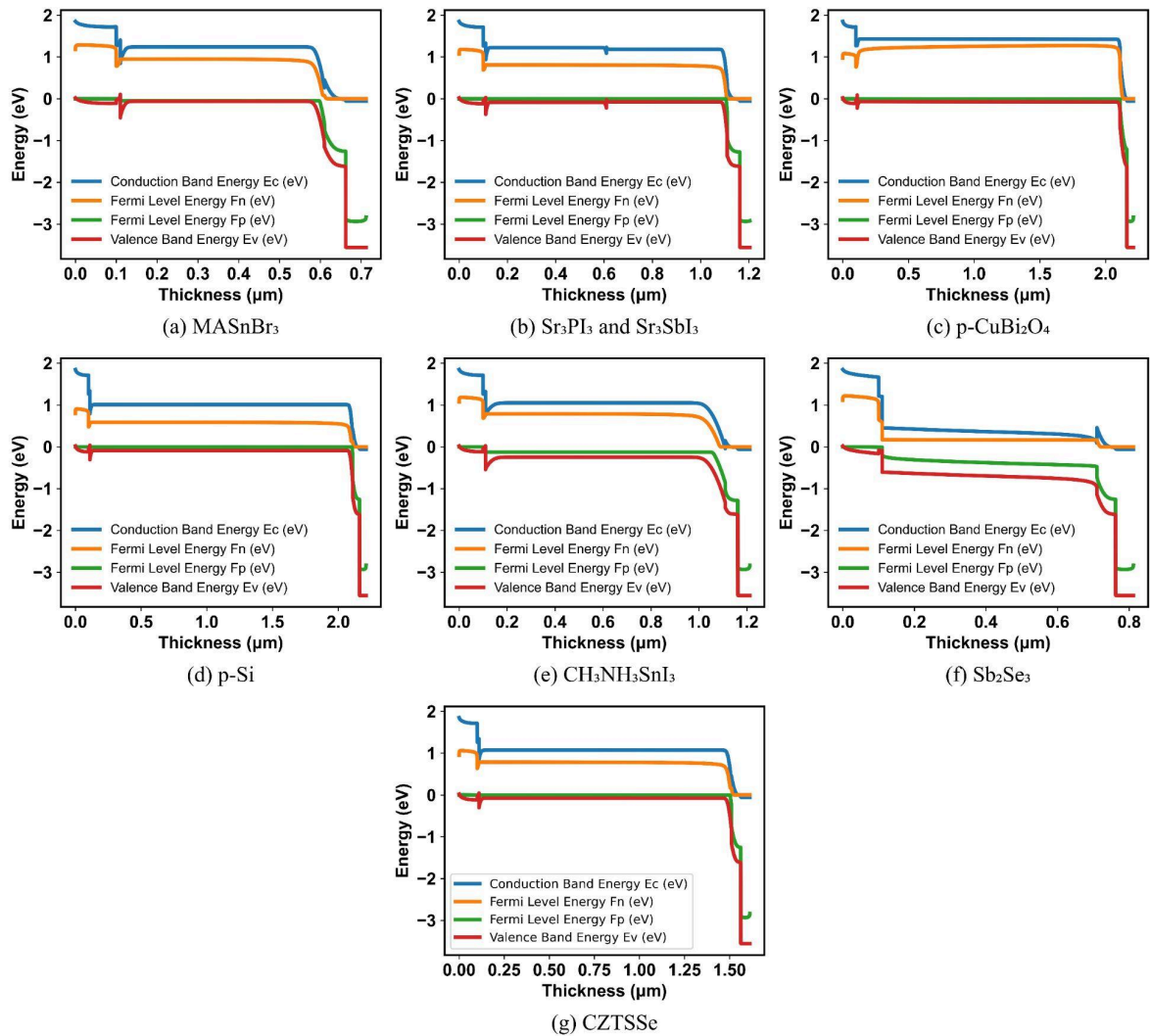

**Fig. 2** Energy band diagrams showing conduction band energy ($E_c$), valence band energy ($E_v$), and Fermi levels for electrons ($F_n$) and holes ($F_p$) across different absorber layer thicknesses.

### 3.2. Current-Voltage and Spectral Response Analysis

Following the energy band analysis, the simulated electrical and optical performance parameters provide further insight into how various absorbers affect the behavior of solar cells under illumination. **Figs. 3** and **4** present the simulated electrical and spectral characteristics of solar cells featuring different absorber materials. **Fig. 3** illustrates the current density-voltage (J-V) behavior, where p-Si demonstrates the highest short-circuit current density of -43.6 mA/cm² due to efficient charge extraction. Conversely, p-CuBi$_2$O$_4$ achieves the largest open-circuit voltage of

1.34 V, attributed to its wider band gap. Materials such as CZTSSe and $Sb_2Se_3$ exhibit competitive current outputs ranging from –41.9 to –42.8 mA/cm² along with moderate voltages of 1.01 V and 1.13 V, respectively, which suggests balanced carrier dynamics. $CH_3NH_3SnI_3$ shows abrupt J-V transitions near 1.0 V, indicating interfacial anomalies.

**Fig. 4** shows the quantum efficiency (QE) spectra, where p-Si and CZTSSe maintain above 90% QE from 300 nm to 1000 nm, enabling broad photon harvesting. $MASnBr_3$ retains high QE up to 900 nm of 86.3% but declines at longer wavelengths, reflecting limited infrared absorption. p-$CuBi_2O_4$ exhibits a narrow spectral response with a QE of approximately 0% beyond 800 nm due to its higher band gap. The integrated analysis indicates that p-Si and CZTSSe are optimal for high-current applications, while $MASnBr_3$ is better suited for voltage-centric designs.

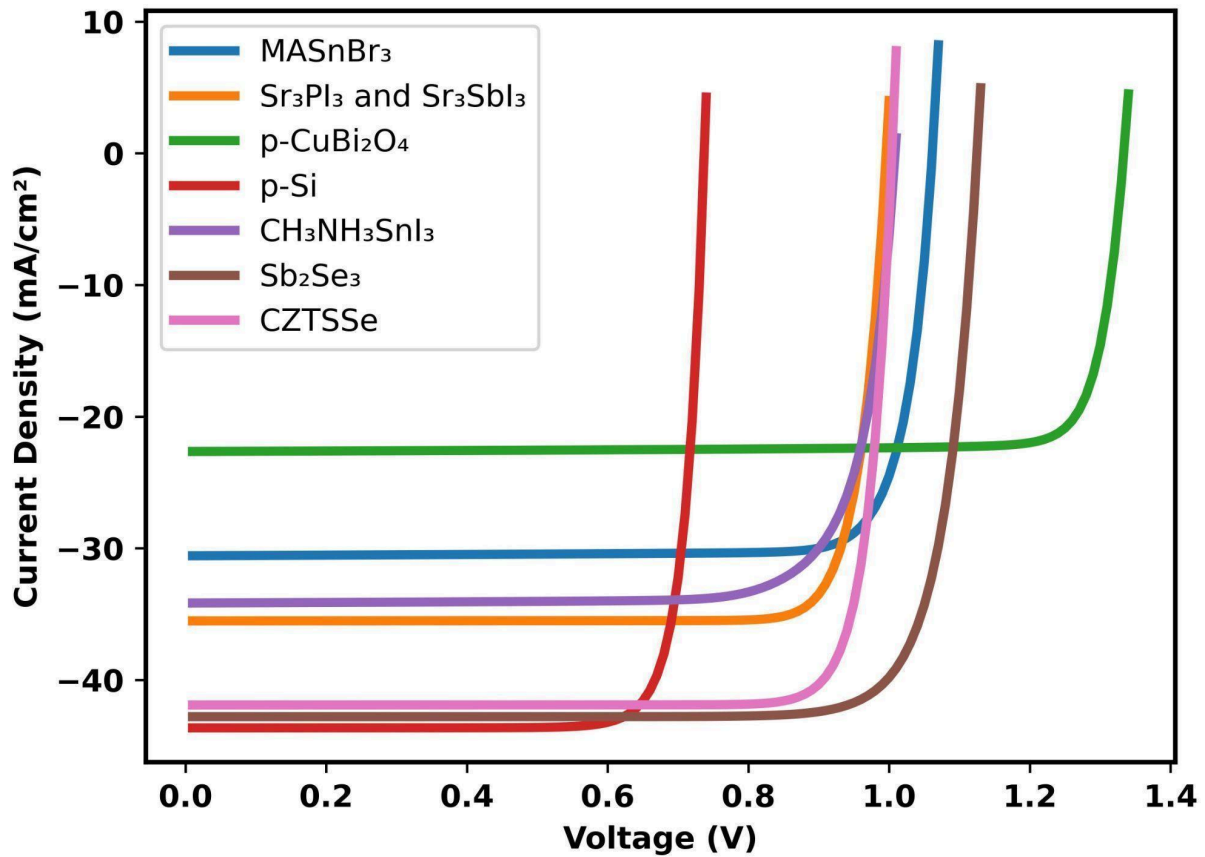

**Fig. 3** Current density-voltage curves for different absorbers under pre-optimized conditions.

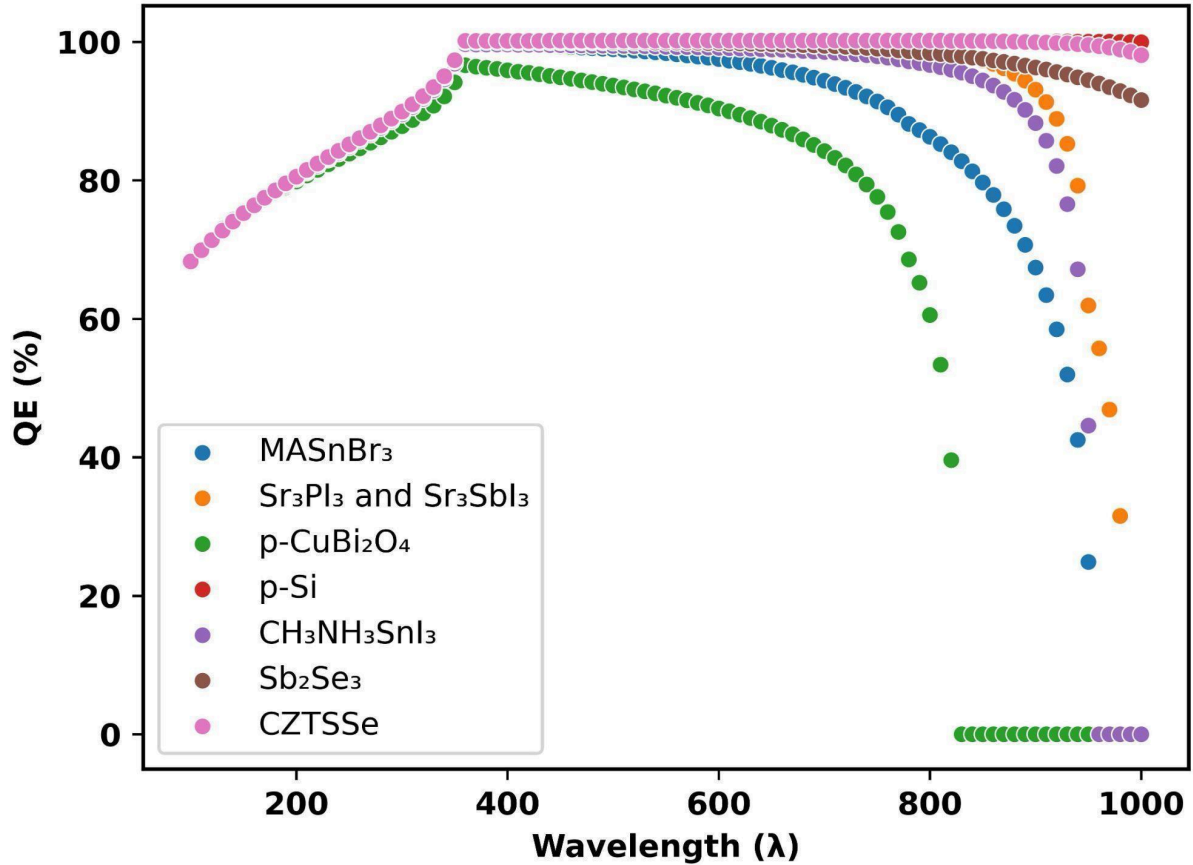

**Fig. 4** Quantum efficiency versus wavelength profiles of different absorbers in the simulated structure.

### 3.3. Photovoltaic Performance Analysis

To comprehensively evaluate the suitability of each absorber, the pre-optimization parameters, including short-circuit current (Jsc), open-circuit voltage (Voc), fill factor (FF), and power conversion efficiency (PCE), are analyzed in detail. **Fig. 5** outlines the pre-optimization parameters, as detailed in **Tables 1-3,** for the seven absorbers integrated into the proposed photovoltaic architecture. Among these, p-CuBi$_2$O$_4$ achieves the highest V$_{oc}$ of 1.33 V, while p-Si exhibits only 0.74 V despite its superior J$_{sc}$ of 43.63 mA/cm$^2$. Notably, both Sb$_2$Se$_3$ and CZTSSe yield J$_{sc}$ values exceeding 41 mA/cm$^2$ due to robust photon absorption. In terms of performance metrics, p-CuBi$_2$O$_4$ also excels in fill factor at 87.81%, in contrast to the 79.76% observed in CH$_3$NH$_3$SnI$_3$, suggesting charge extraction inefficiencies. The material Sb$_2$Se$_3$ dominates PCE of 39.99% by balancing J$_{sc}$ of 42.78 mA/cm$^2$ and V$_{oc}$ of 1.0 V, while CZTSSe follows with 36.31%

PCE through high FF of 86.31%. The relatively low $V_{oc}$ of p-Si limits its overall PCE despite its strong $J_{sc}$ performance. These results underscore the intrinsic trade-offs associated with different absorber materials: oxide-based absorbers tend to favor voltage output, whereas chalcogenides are more conducive to maximizing current density. The pre-optimization data highlight $Sb_2Se_3$ and CZTSSe as frontrunners for further development in high-efficiency photovoltaics.

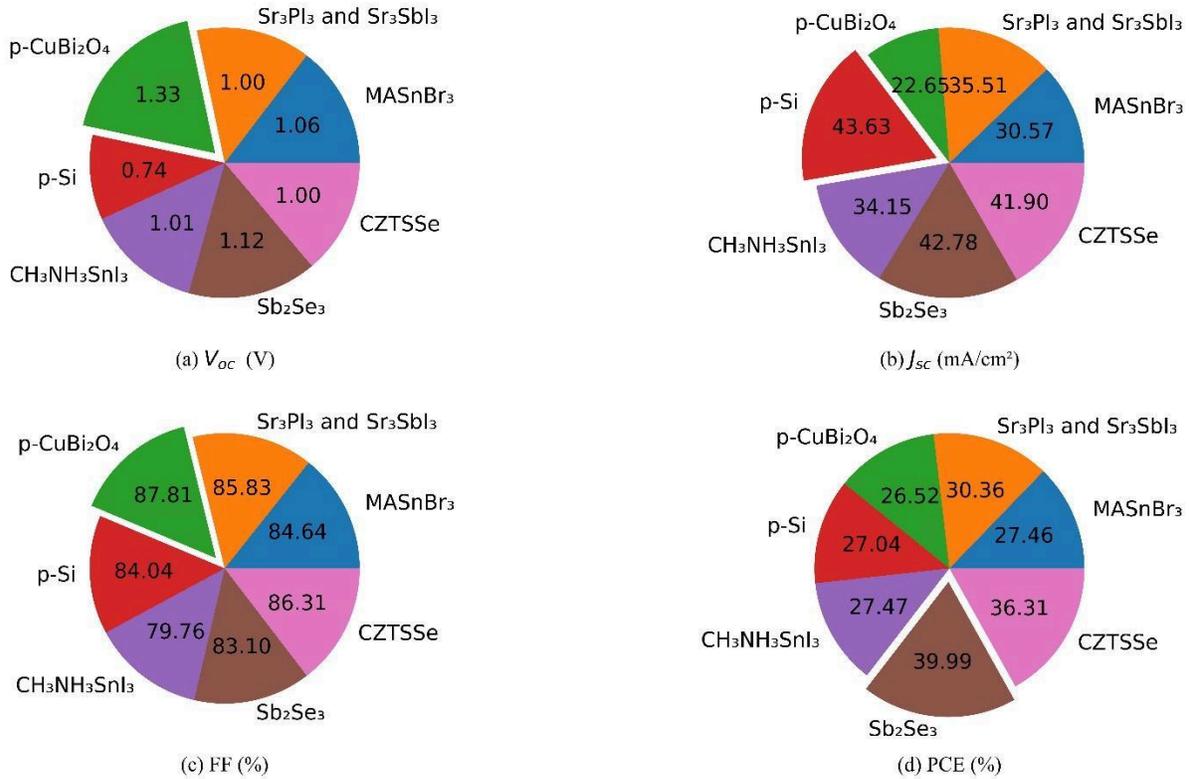

**Fig. 5** Initial performance parameters (a) open-circuit voltage ($V_{oc}$), (b) short-circuit current density ($J_{sc}$), (c) fill factor (FF), and (d) power conversion efficiency (PCE) for various absorber materials in the structure.

### 3.4. Parametric Analysis of Absorber Materials

#### 3.4.1. Impact of Thickness

Understanding how absorber layer thickness influences charge generation and transport is crucial for achieving optimal device performance. **Fig. 6** presents a comprehensive analysis of how the thickness of absorber materials impacts various performance parameters across seven distinct materials. $MASnBr_3$, with a thickness range of 0.5 to 3 μm **(Fig. 6a)**, exhibits gradual efficiency gains from 27% to 30% with an optimal PCE of 29.98% at 2.2 μm. Similarly, $Sr_3PI_3$ and $Sr_3SbI_3$ **(Fig. 6b-c)** achieve peak efficiencies of 31.49% and 30.77% near 2.2 μm and 2.0 μm, respectively. Both materials exhibit a stable $V_{oc}$ of approximately 0.996 V, indicating that the voltage output is relatively insensitive to variations in thickness. In contrast, p-$CuBi_2O_4$ **(Fig. 6d)** maintains consistent performance with 26.52% PCE across the 1.9 to 2.5 μm range due to minimal shifts in photovoltaic parameters. p-Si **(Fig. 6e)** reaches peak efficiency at 1.2 μm with 27.21% PCE but shows signs of recombination losses beyond this thickness. $CH_3NH_3SnI_3$ **(Fig. 6f)** performs best with thinner layers around 0.9 μm, achieving 27.49% PCE by effectively balancing light absorption and charge extraction. $Sb_2Se_3$ **(Fig. 6g)** performs best at 0.9 μm with a high PCE of 40.43%, supported by a strong $J_{sc}$ of 43.89 mA/cm² and $V_{oc}$ of 1.12 V, which demonstrates highly efficient thin-film operation. Finally, CZTSSe **(Fig. 6h)** reaches 36.46% PCE at 1.9 μm through balanced $V_{oc}$-$J_{sc}$ trade-offs. These findings reveal material-specific thickness thresholds at which light harvesting and carrier transport are optimally aligned. Robust performers like $Sr_3PI_3$ and $Sb_2Se_3$ enable efficient thin devices, whereas other materials require precise thickness control to mitigate losses.

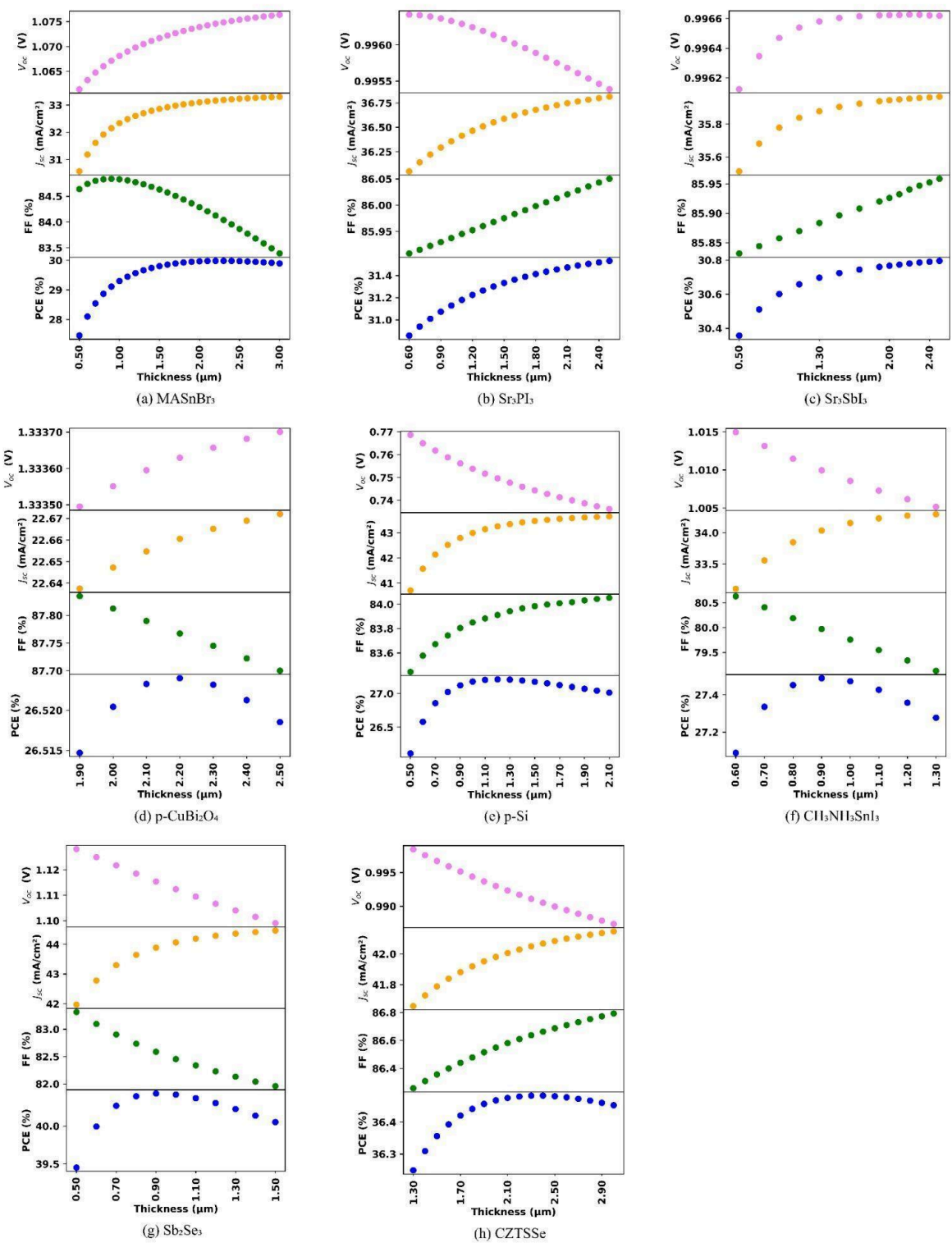

**Fig. 6** Influence of absorber layer thickness on photovoltaic parameters ($V_{oc}$, $J_{sc}$, FF, and PCE) for different absorber materials.

### 3.4.2. Impact of Doping Density

Doping density has a significant impact on charge carrier concentration and recombination rates; therefore, analyzing its influence is crucial for determining the optimal doping levels for each material. Fig. 7 maps the correlation between doping density and the performance parameters of various proposed absorber materials. In particular, $MASnBr_3$ (Fig. 7b) exhibits donor-driven gains in $V_{oc}$ of 1.13 V and FF of 84.36% peaking at $10^{18}$ cm$^{-3}$. In contrast, $Sr_3PI_3$ and $Sr_3SbI_3$ (Fig. 7c-d) show minimal fluctuation in optimum PCE values of 35.97% and 34.59%, respectively, across acceptor densities ranging from $10^{19}$ to $10^{22}$ cm$^{-3}$. This observed stability in PCE suggests a defect-tolerant band alignment inherent to these materials. Additionally, p-$CuBi_2O_4$ achieves a maximal $V_{oc}$ of 1.43 V and PCE of 28.65 % at high acceptor doping of $3.7 \times 10^{20}$ cm$^{-3}$ (Fig. 7e) due to enhanced hole mobility. Furthermore, p-Si shows a balance between $J_{sc}$ of 43.25 mA/cm$^2$ at low doping and $V_{oc}$ of 0.75 V at higher doping levels (Fig. 7f). This behavior reflects the inherent trade-offs between carrier concentration and photovoltaic response. $CH_3NH_3SnI_3$ (Fig. 7g) maintains a stable $J_{sc}$ of approximately 32 mA/cm$^2$ and FF above 80% within $10^{17}$ to $10^{19}$ cm$^{-3}$ but peaks at 28.88% PCE. Conversely, $Sb_2Se_3$ (Fig. 7h) excels with a 40.4% PCE at an ultralow acceptor density of $2 \times 10^{13}$ cm$^{-3}$, underscoring its intrinsic defect resilience. Lastly, CZTSSe (Fig. 7i) requires high doping of $10^{21}$ cm$^{-3}$ for optimal conductivity and 40%, which can be attributed to its complex stoichiometric properties.

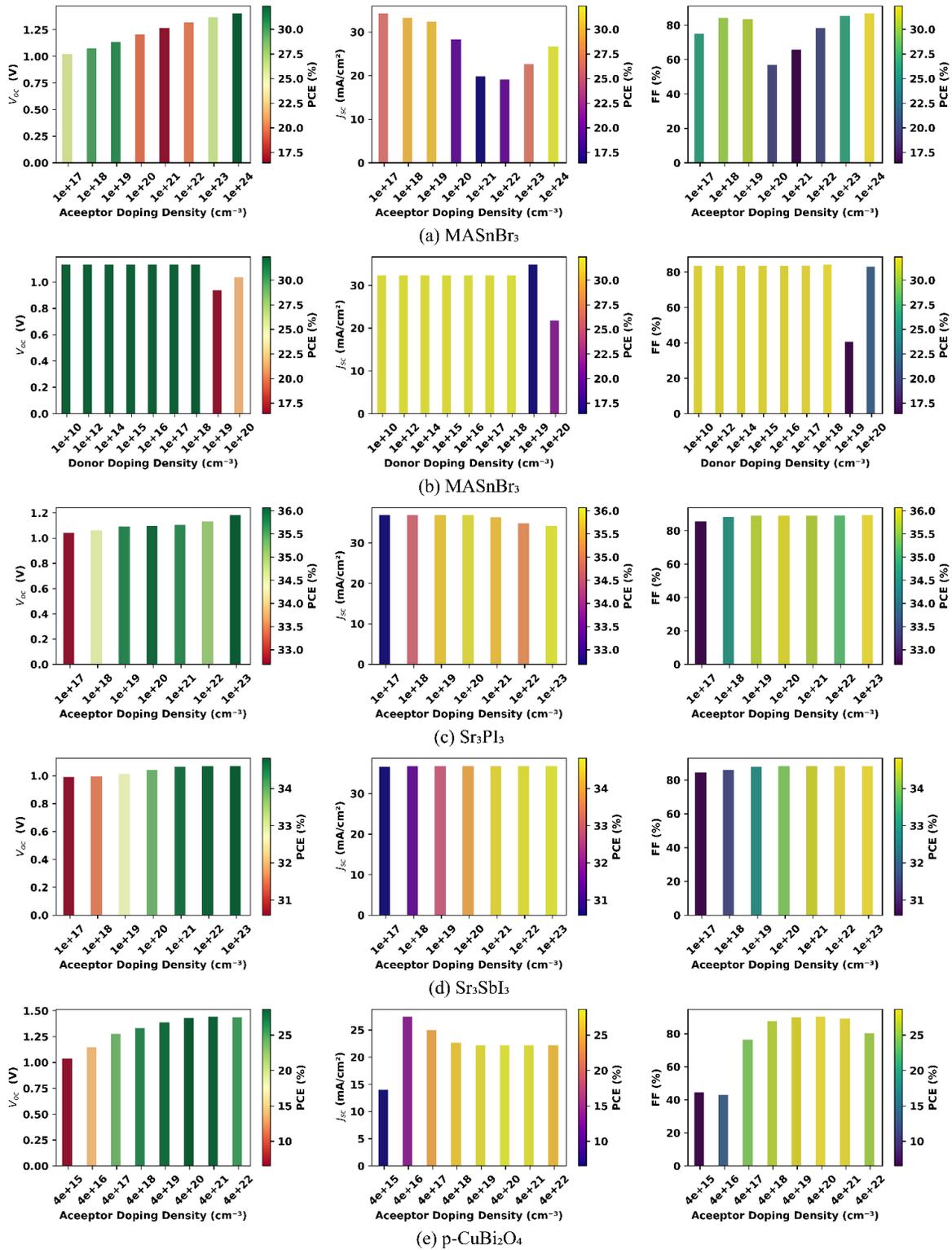

**Fig. 7** Effect of acceptor doping density ($N_a$) and donor doping density ($N_d$) on photovoltaic parameters ($V_{oc}$, $J_{sc}$, FF, and PCE) for different absorber materials.

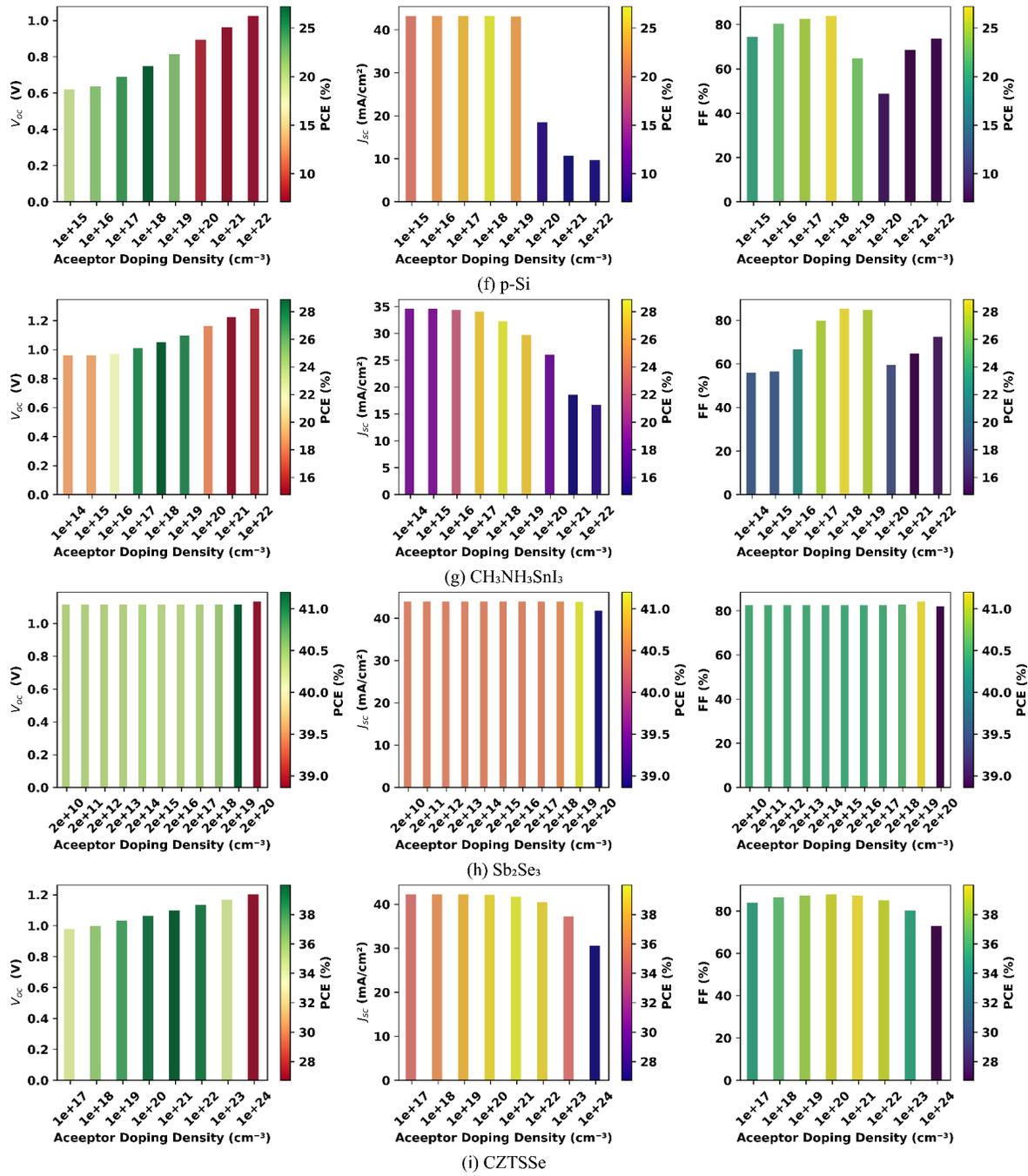

**Fig. 7** (Continued).

### 3.4.3. Impact of Defect Density

For solar cells, the defect density has a significant impact on photovoltaic performance across all absorber materials but with varying sensitivities as shown in **Fig. 8**. For MASnBr$_3$ **(Fig. 8a)**, the performance parameters remain stable until the defect density exceeds $10^{13}$ cm$^{-3}$, beyond which the PCE drops sharply from 32% to 5% with further increases in defect density. Conversely, both Sr-based materials **(Fig. 8b-c)** show remarkable defect tolerance with $Sr_3SbI_3$ maintaining a high PCE of approximately 36% across a wide defect range. The p-$CuBi_2O_4$ absorber **(Fig. 8d)** is acutely sensitive to defects, with PCE plummeting from 34.7% to 9.8% as defect density increases. Additionally, the p-Si material **(Fig. 8e)** demonstrates robust performance up to $2\times10^{10}$ cm$^{-3}$ but deteriorates rapidly beyond this threshold. $CH_3NH_3SnI_3$ **(Fig. 8f)** shows moderate defect tolerance with a gradual decline in performance above $10^{12}$ cm$^{-3}$. Notably, $Sb_2Se_3$ **(Fig. 8g)** maintains exceptional performance even at higher defect densities with PCE above 40% at $10^5$-$10^{10}$ cm$^{-3}$. CZTSSe **(Fig. 8h)** shows good tolerance up to $10^{12}$ cm$^{-3}$ before significant performance loss occurs. Across all materials analyzed, $V_{oc}$ is most affected by increasing defect density due to enhanced recombination mechanisms. Conversely, $J_{sc}$ also decreases with higher defect concentrations, but less dramatically than $V_{oc}$ in most cases. These findings highlight the importance of developing defect management strategies tailored to specific absorber materials. The results suggest that $Sb_2Se_3$ and $Sr_3SbI_3$ offer superior defect tolerance, presenting significant advantages for the fabrication of practical photovoltaic devices, especially in contexts where defect control poses substantial challenges.

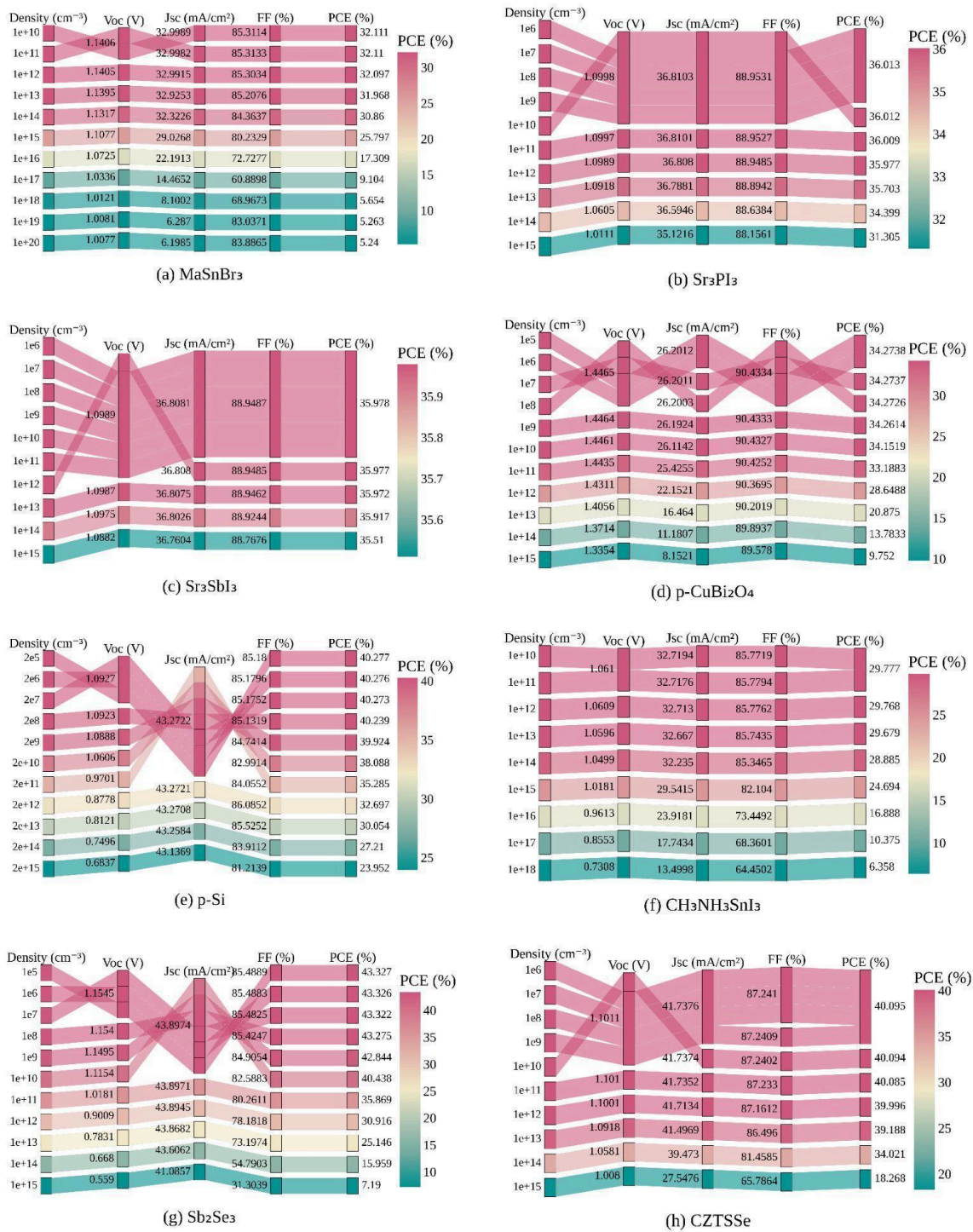

**Fig. 8** Effect of defect density variation on the photovoltaic parameters ($V_{oc}$, $J_{sc}$, FF, and PCE) of different absorber materials.

### 3.4.4. Impact of Temperature

Temperature is a critical environmental factor that influences charge recombination, carrier mobility, and overall photovoltaic stability. Assessing thermal behavior helps determine the suitability of materials for diverse operational conditions. **Fig. 9** represents the photovoltaic performance of seven absorber materials for a temperature variation from 280 to 350 K. p-$CuBi_2O_4$ maintains exceptional $V_{oc}$ stability at approximately 1.4 V and minimal PCE loss of less than 3% up to 350 K, demonstrating superior high-temperature resilience. In contrast, CZTSSe peaks near 41.18% PCE at 280 K but degrades by 10.4% at 350 K due to accelerated recombination, favoring cooler operating conditions. Furthermore, $CH_3NH_3SnI_3$ shows a severe $V_{oc}$ collapse from 1.1 to 0.9 V and a PCE decline from 31.87% to 25.65%, with optimal performance restricted to 280–300 K. On the other hand, p-Si sustains a $J_{sc}$ of 43.27 mA/cm² across a wide range of temperatures but performs best below 310 K to mitigate $V_{oc}$ losses. The materials $Sr_3PI_3$/$Sr_3SbI_3$ retain a PCE range between 33.45% and 36.96% across the studied range. Notably, the fill factor drops uniformly for p-$CuBi_2O_4$, which resists the decline from 90.97% to 88.17%, while the others also degrade rapidly, except for $MASnBr_3$. Therefore, p-$CuBi_2O_4$ emerges as a promising candidate for high-temperature applications, whereas $Sb_2Se_3$ and CZTSSe excel in moderate temperatures. So, the selection of absorbers for specific environmental uses is driven by their temperature dependency. However, all absorber materials here exhibit their optimum photovoltaic performance at 280 K.

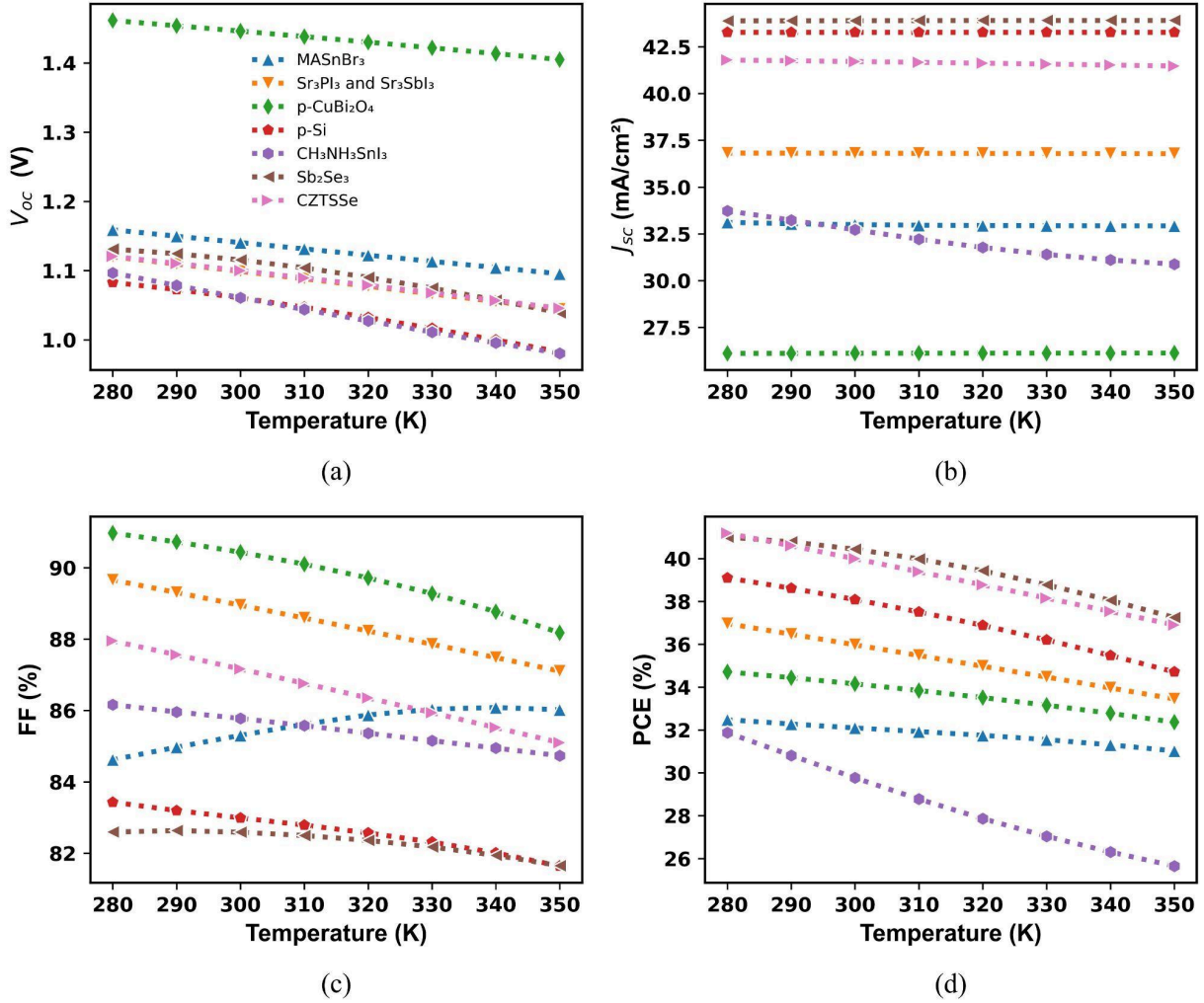

**Fig. 9** Temperature dependence of photovoltaic parameters with different absorber materials: (a) open-circuit voltage ($V_{oc}$), (b) short-circuit current density ($J_{sc}$), (c) fill factor (FF), and (d) power conversion efficiency (PCE).

### 3.5. Final Optimization Results

To improve the photovoltaic performance of the proposed device architectures, a systematic optimization of the key structural and electrical parameters was conducted, as outlined in **Table 4**. For MASnBr$_3$, the absorber thickness increased from 0.5 μm to 2.2 μm, which coincided with a tenfold increase in acceptor density from $10^{18}$ to $10^{19}$ cm$^{-3}$ and alongside a reduction in defect density from $10^{14}$ to $10^{12}$ cm$^{-3}$. Both Sr$_3$SbI$_3$ and Sr$_3$PI$_3$ exhibited substantial increases in acceptor concentrations, reaching values of $10^{21}$ cm$^{-3}$ and $10^{20}$ cm$^{-3}$, respectively. The defect densities for these materials were taken from $10^{14}$ cm$^{-3}$ to $10^{12}$ cm$^{-3}$. The defect density of the p-CuBi$_2$O$_4$ layer was varied by two orders of magnitude from $10^{12}$ to $10^{10}$ cm$^{-3}$, while p-Si achieved a notable defect suppression from $2\times10^{14}$ to $2\times10^{10}$ cm$^{-3}$. Additionally, Sb$_2$Se$_3$ and CZTSSe showed optimized thicknesses of 0.9 μm and 1.9 μm, respectively. Thermal tuning to 280 K further enhanced the stability and carrier dynamics across all absorber materials.

Table 4 Optimized parameters for the initial structure of the cell.

| Parameters | Thickness (μm) | Acceptor Density $N_a$ | Donor Density $N_d$ | Total Defect Density $N_t$ | Temperature (K) |
|---|---|---|---|---|---|
| MASnBr$_3$ | 2.2 | 1.00E+19 | 1.00E+18 | 1.00E+12 | 280 |
| Sr$_3$PI$_3$/Sr$_3$SbI$_3$ | Sr$_3$SbI$_3$–2, Sr$_3$PI$_3$–2.2 | Sr$_3$SbI$_3$–1.00E+21, Sr$_3$PI$_3$–1.00E+20 | 0 | Sr$_3$SbI$_3$–1.00E+12, Sr$_3$PI$_3$–1.00E+12 | |
| p-CuBi$_2$O$_4$ | 2.2 | 3.70E+20 | 0 | 1.00E+10 | |
| p-Si | 1.2 | 1.00E+18 | 0 | 2.00E+10 | |
| CH$_3$NH$_3$SnI$_3$ | 0.9 | 1.00E+18 | 0 | 1.00E+12 | |
| Sb$_2$Se$_3$ | 0.9 | 2.00E+13 | 0 | 1.00E+10 | |
| CZTSSe | 1.9 | 1.00E+21 | 0 | 1.00E+12 | |

This systematic optimization yielded significant improvements in the performance parameters, as shown in **Table 5**. Among the materials examined, CZTSSe achieved the highest PCE of 41.19%, characterized by a $J_{sc}$ of 41.79 mA/cm$^2$ and a FF of 87.95%. Sb$_2$Se$_3$ closely followed at 41% PCE, driven by its exceptional $J_{sc}$ of 43.89 mA/cm$^2$. The p-Si absorber attained a PCE of 39.1%, supported by a balanced $J_{sc}$ of 43.27 mA/cm$^2$ and a $V_{oc}$ of 1.083 V. The Sr$_3$PI$_3$/Sr$_3$SbI$_3$ combination demonstrated a superior FF of 89.65%, contributing to its 36.96% PCE. Moderate efficiencies were recorded for MASnBr$_3$ and CH$_3$NH$_3$SnI$_3$ at 32.48% and 31.87%, respectively, primarily limited by their lower $J_{sc}$ values. Notably, p-CuBi$_2$O$_4$ achieved the highest $V_{oc}$ of 1.461 V; however, its PCE of 34.71% was constrained by a comparatively modest $J_{sc}$ of 26.11 mA/cm$^2$.

Table 5 Optimized simulation results using different absorber materials.

| List of Absorbers | $V_{oc}$ (V) | $J_{sc}$ (mA/cm$^2$) | FF (%) | PCE (%) |
|---|---|---|---|---|
| MASnBr$_3$ | 1.1591 | 33.116727 | 84.62 | 32.48 |
| Sr$_3$PI$_3$/Sr$_3$SbI$_3$ | 1.1198 | 36.817328 | 89.65 | 36.96 |
| p-CuBi$_2$O$_4$ | 1.4616 | 26.109243 | 90.97 | 34.71 |
| p-Si | 1.083 | 43.272068 | 83.43 | 39.1 |
| $CH_3NH_3SnI_3$ | 1.0968 | 33.729164 | 86.16 | 31.87 |
| $Sb_2Se_3$ | 1.1311 | 43.887387 | 82.59 | 41 |
| CZTSSe | 1.1206 | 41.792132 | 87.95 | 41.19 |

Following the optimization of performance parameters, the study further investigated the compatibility of various transparent conducting oxide (TCO) substrates as potential alternatives to FTO. **Table 6** presents the simulation results using ITO, IZO, and MZO [87] as transparent conducting substrates. The analysis encompassed seven absorbers with identical structural configurations and all optimized parameters. Among the alternatives, ITO demonstrated limited compatibility. It performed acceptably only with the $MASnBr_3$ absorber, achieving a FF of 56.41% and a PCE of 22.76%. Conversely, IZO was found to lack compatibility with any of the selected absorber materials within this structural configuration. In stark contrast, MZO exhibited promising results. It was compatible with all seven absorbers, exhibiting only a marginal reduction in PCE, quantified at 0.17% to 2.42% compared to the benchmark FTO. This suggests that MZO could serve as a viable alternative to FTO in applications requiring transparent conducting substrates.

**Table 6** Simulation results of transparent conducting substrates with seven optimized absorbers.

| Structure | $V_{oc}$ (V) | $J_{sc}$ (mA/cm$^2$) | FF (%) | PCE (%) |
|---|---|---|---|---|
| ITO/$PeDAMA_5Pb_6I_{19}$/IDL1/$MASnBr_3$/IDL2/$PeDAMA_2Pb_3I_{10}$/C | 1.22 | 33.076546 | 56.41 | 22.76 |
| MZO/ $PeDAMA_5Pb_6I_{19}$/IDL1/$MASnBr_3$/IDL2/ $PeDAMA_2Pb_3I_{10}$/C | 1.159 | 33.100481 | 84.36 | 32.36 |
| MZO/ $PeDAMA_5Pb_6I_{19}$/IDL1/double absorber ($Sr_3PI_3$/$Sr_3SbI_3$)/IDL2/ $PeDAMA_2Pb_3I_{10}$/C | 1.1197 | 36.800916 | 89.45 | 36.86 |
| MZO/ $PeDAMA_5Pb_6I_{19}$/IDL1/p-$CuBi_2O_4$/IDL2/ $PeDAMA_2Pb_3I_{10}$/C | 1.443 | 26.10069 | 89.93 | 33.87 |
| MZO/ $PeDAMA_5Pb_6I_{19}$/IDL1/p-Si/IDL2/ $PeDAMA_2Pb_3I_{10}$/C | 1.0828 | 43.226531 | 83.35 | 39.02 |
| MZO/ $PeDAMA_5Pb_6I_{19}$/IDL1/$CH_3NH_3SnI_3$/IDL2/ $PeDAMA_2Pb_3I_{10}$/C | 1.0965 | 33.683577 | 86.05 | 31.78 |
| MZO/ $PeDAMA_5Pb_6I_{19}$/IDL1/$Sb_2Se_3$/IDL2/ $PeDAMA_2Pb_3I_{10}$/C | 1.1309 | 43.829462 | 82.58 | 40.93 |
| MZO/ $PeDAMA_5Pb_6I_{19}$/IDL1/CZTSSe/IDL2/ $PeDAMA_2Pb_3I_{10}$/C | 1.1205 | 41.725853 | 87.94 | 41.11 |

**Table 7** lists a comparative analysis of various absorber materials incorporated into different device architectures and highlights their corresponding power conversion efficiency. This study identifies CZTSSe and $Sb_2Se_3$ as high-efficiency absorbers for perovskite-inspired solar cells, with PCEs exceeding 41%. It is evident that the systematic optimization of both interfacial and bulk properties has markedly improved carrier extraction and minimized loss mechanisms, thus contributing to enhanced device performance.

Table 7 Comparison of different structures.

| Structure | PCE (%) |
|---|---|
| FTO/SnO$_2$/MASnBr$_3$/NiO/Au | 34.52 [85] |
| Al/FTO/ZnSe/double absorber (Sr$_3$PI$_3$/Sr$_3$SbI$_3$)/Au | 34.13 [63] |
| Al/FTO/CdS/CuBi$_2$O$_4$/Ni | 26 [81] |
| ZnO:Al/CdS/CZTS/Si/Mo | 19.4 [82] |
| FTO/TiO$_2$/CH$_3$NH$_3$SnI$_3$/Ni/glass | 26.33 [83] |
| TO/IGZO/CdS/Sb$_2$Se$_3$/PbS/Au | 36.11 [84] |
| FTO/STO/CsPbI$_3$/CZTSSe/NiO/W | 37.35 [30] |
| FTO/TiO$_2$/IDL1/PeDAMA$_5$Pb$_6$I$_{19}$/MAPbI$_3$/IDL2/Ni | 29.37 [80] |
| FTO/ PeDAMA$_5$Pb$_6$I$_{19}$/IDL1/CsSnI3/IDL2/PeDAMA$_2$Pb$_3$I$_{10}$/C | 37.83 [80] |
| FTO/ PeDAMA$_5$Pb$_6$I$_{19}$/IDL1/MASnBr$_3$/IDL2/ PeDAMA$_2$Pb$_3$I$_{10}$/C | 32.48 (Present Study) |
| FTO/ PeDAMA$_5$Pb$_6$I$_{19}$/IDL1/double absorber (Sr$_3$PI$_3$/Sr$_3$SbI$_3$)/IDL2/ PeDAMA$_2$Pb$_3$I$_{10}$/C | 36.96 (Present Study) |
| FTO/ PeDAMA$_5$Pb$_6$I$_{19}$/IDL1/p-CuBi$_2$O$_4$/IDL2/ PeDAMA$_2$Pb$_3$I$_{10}$/C | 34.71 (Present Study) |
| FTO/ PeDAMA$_5$Pb$_6$I$_{19}$/IDL1/p-Si/IDL2/ PeDAMA$_2$Pb$_3$I$_{10}$/C | 39.1 (Present Study) |
| FTO/ PeDAMA$_5$Pb$_6$I$_{19}$/IDL1/CH$_3$NH$_3$SnI$_3$/IDL2/ PeDAMA$_2$Pb$_3$I$_{10}$/C | 31.87 (Present Study) |
| FTO/ PeDAMA$_5$Pb$_6$I$_{19}$/IDL1/Sb$_2$Se$_3$/IDL2/ PeDAMA$_2$Pb$_3$I$_{10}$/C | 41 (Present Study) |
| FTO/ PeDAMA$_5$Pb$_6$I$_{19}$/IDL1/CZTSSe/IDL2/ PeDAMA$_2$Pb$_3$I$_{10}$/C | 41.19 (Present Study) |

4. **Conclusion**

This research has systematically explored high-efficiency perovskite solar cells of seven lead-free absorber materials. The performance was analyzed within a consistent device architecture to examine the impact of the layer thickness, doping concentration, defect density, and temperature. The results indicate that the material choice has a significant effect on device efficiency. Our findings reveal that CZTSSe and Sb$_2$Se$_3$ stand out as exceptional candidates, achieving remarkable efficiencies of 41.19% and 41%, respectively. The optimized 2D/3D/2D architecture demonstrates that band alignment and defect suppression enhance charge extraction while minimizing recombination. This study highlights that the absorber material thickness and doping thresholds critically influence the performance parameters of PSCs, where $Sb_2Se_3$ achieves optimal efficiency at 0.9 μm and CZTSSe requires 1.9 μm for balanced carrier transport. Here, p-$CuBi_2O_4$ demonstrated excellent voltage performance and thermal resilience, retaining 97% efficiency at 350 K. In contrast, $Sb_2Se_3$ and CZTSSe show high efficiency in moderate temperatures but degrade under extreme heat. Additionally, defect tolerance studies have highlighted Sr-based materials as robust candidates for scalable production, maintaining approximately 36% efficiency across a wide range of defect densities. Similarly, p-Si demonstrates strong current generation capabilities, although its voltage output remains a limiting factor for achieving higher overall efficiency. To address these limitations, optimizing doping profiles and defect management strategies was found to further enhance device performance. Notably, materials such as $MASnBr_3$ and p-$CuBi_2O_4$ exhibited high sensitivity to defect levels, with low defect densities being crucial to prevent rapid efficiency degradation. In parallel, the study evaluated the potential of alternative transparent conducting oxides. Among them, MZO proved to be a viable substitute for FTO, offering broad compatibility without significant performance loss. Conversely, ITO and IZO are less effective for the given architecture. While this study provides valuable insights through simulation, integrating non-perovskite absorbers into perovskite-based stacks remains a significant challenge due to bandgap mismatches and structural incompatibilities. Therefore, future research should focus on experimentally validating these findings and investigating long-term stability and scalability.

## CRediT authorship contribution statement

**Md. Meraz Hasan:** Writing – original draft, Writing – review & editing, Conceptualization, Methodology, Software, Formal analysis, Visualization, Data curation. **Pallab Chakraborty:** Writing – review & editing, Methodology, Formal analysis, Visualization, Data curation, Validation. **Fahim Tanvir:** Writing – review & editing, Formal analysis, Visualization. **Subah Tahsin:** Writing – original draft, Formal analysis. **Mostafizur Rahaman:** Resources, Supervision, Conceptualization, Writing – original draft, Writing – review & editing,, Methodology, Visualization, Formal analysis.

## Declaration of competing interest

We declare no conflict of interest.

**Data availability**

The data that support the findings of this study are available from the corresponding author upon reasonable request.

**Acknowledgements**

The authors acknowledge and are grateful to Dr. Marc Burgelman of the University of Gent, Belgium, for providing the SCAPS 1-D simulation software. The authors acknowledge everyone who contributed to this research.

**Declaration of generative AI and AI-assisted technologies in the writing process**

During the preparation of this work, the authors used ChatGPT 3.5, developed by OpenAI, to refine, edit, and condense sections of this paper during the writing process. After using this tool/service, the authors reviewed and edited the content as needed and take full responsibility for the content of the publication.